\documentclass[12pt]{article} 
\usepackage{amssymb,latexsym}
\def\beq{\begin{equation}}
\def\eeq{\end{equation}}
\def\bea{\begin{eqnarray}}
\def\eea{\end{eqnarray}}
\def\beas{\begin{eqnarray*}}
\def\eeas{\end{eqnarray*}}
\def\nn{\nonumber}
\def\ba{\begin{array}}
\def\ea{\end{array}}
\def\al{\alpha}
\def\be{\beta}
\def\ga{\gamma}
\def\la{\lambda}
\def\th{\theta}
\def\vp{\varphi}
\newtheorem{theo}{Theorem}

\newtheorem{lemm}[theo]{Lemma}
\newtheorem{coro}[theo]{Corollary}

\newtheorem{prop}[theo]{Proposition}
\def\Nat{\mathbb{N}}
\def\Zah{\mathbb{Z}}
\def\Real{\mathbb{R}}
\def\C{\mathbb{C}}
\begin{document}
\renewcommand{\thefootnote}{\fnsymbol{footnote}}
\setcounter{footnote}{1}
\begin{flushright}
math-ph/9807019
\end{flushright}
\begin{center}
{\Large \bf  Realizations of $su(1,1)$ and $U_q(su(1,1))$ }\\[2mm]
{\Large \bf and generating functions for orthogonal polynomials}

\vspace{2cm}
J.\ Van der Jeugt\footnote{Research Associate of the
Fund for Scientific Research -- Flanders (Belgium). 
E-mail address~: Joris.VanderJeugt@rug.ac.be.} and R.\
Jagannathan\footnote{Permanent address~: Institute of Mathematical
Sciences, CIT Campus, Madras 600113, India. E-mail address~:
jagan@imsc.ernet.in.}

\vspace{0.5cm}
Department of Applied Mathematics and Computer Science \\ 
University of Ghent, Krijgslaan 281-S9, B-9000 Gent, Belgium 
\end{center}

\vspace{2cm}
\noindent
{\bf Abstract}:  
Positive discrete series representations of the Lie algebra $su(1,1)$
and the quantum algebra $U_q(su(1,1))$ are considered. 
The diagonalization of a self-adjoint operator (the Hamiltonian) in
these representations and in tensor products of such representations is
determined, and the generalized eigenvectors are constructed in terms
of orthogonal polynomials. 
Using simple realizations of $su(1,1)$, $U_q(su(1,1))$, and their
representations, these generalized eigenvectors are shown to coincide
with generating functions for orthogonal polynomials. 
The relations valid in the tensor product representations then give
rise to new generating functions for orthogonal polynomials, or to
Poisson kernels. In particular, a group theoretical derivation of the
Poisson kernel for Meixner-Pollaczak and Al-Salam--Chihara polynomials
is obtained.

\vspace{1cm}

\noindent
PACS~: 02.20.+b, 02.30.+g, 03.65.Fd.

\vspace{1cm}

\noindent
to appear in {\em J.\ Math.\ Phys.}

\newpage
\renewcommand{\thesection}{\Roman{section}}
\renewcommand{\theequation}{\arabic{section}.{\arabic{equation}}}
\setcounter{equation}{0}
\section{Introduction}

The Lie algebra $su(1,1)$ (or $so(2,1)$) has been extensively used as
spectrum generating algebra in many simple quantum systems, such as the
non-relativistic Coulomb problem, the isotropic harmonic oscillator,
Schr\"odinger's relativistic equation, and the Dirac-Coulomb
problem~\cite[and references therein]{Adams}. In some interacting boson
models the Hamiltonian $H$ can be written as a linear combination of
the $su(1,1)$ basis elements $J_0$, $J_{\pm}$, such that $H^\dagger
=H$, see e.g.\ the superfluid bose system~\cite{Solomon,Gerry}. In such
boson models, the representations that play a role are the positive
discrete series representations of $su(1,1)$, often denoted by ${\cal
D}^+(k)$. For a given Hamiltonian $H=a J_0 + b J_+ + c J_- +d$, with
$a^*=a$, $d^*=d$ and $b^*=c$, the important questions are~: (1) its
spectrum (continuous or discrete); (2) the expansion of its
eigenvectors in terms of the boson states $|k,n\rangle$ (corresponding
to eigenstates of $J_0$ in the representations ${\cal D}^+(k)$).

In~\cite{Vanderjeugt} a particular element, $H=2J_0-J_+-J_-$, was
chosen, and it was shown that the expansion coefficients of the
eigenstates of $H$ in the basis $|k,n\rangle$ of the representation
${\cal D}^+(k)$ are given as generalized Laguerre polynomials (see
also~\cite{GZ}). The main 
purpose of~\cite{Vanderjeugt} was then to consider tensor product
representations, and deduce in a Lie algebraic way addition or
convolution formulas for orthogonal polynomials (in this case for
generalized Laguerre and Jacobi polynomials,
cfr.~\cite[(1.1)]{Vanderjeugt}). 
Following this, the most general Hamiltonian with real coefficients
(up to an overall constant
and scaling factor) $H=\sigma J_0-J_+-J_-$ ($\sigma \in \Real$) was
considered in~\cite{KV}, and it was shown that the spectrum of $H$
depends upon whether $|\sigma|=2$, $|\sigma|<2$ or $|\sigma|>2$ (see
also~\cite{Bacry} for a particular representation). In
each of these cases, the expansion coefficients of the Hamiltonian
eigenstates in the $|k,n\rangle$-basis are orthogonal polynomials. 
In section~II, the main results of~\cite{Vanderjeugt,KV} are summarized
in the present framework of eigenstates (generalized eigenvectors). 

Although the physical applications or interpretations are clear from
the above observations, this paper is primarily dealing with a number
of mathematical consequences of choosing particular realizations for
the positive discrete series representations. 

The general framework for the positive discrete series representations
is outlined in section~II. The three type of operators are given,
together with their spectrum and expansion coefficients (orthogonal
polynomials). The tensor product of two positive discrete series
representations is considered, and just as there are ``coupled'' and
``uncoupled'' eigenvectors for $J_0$ in the tensor product, one can
also define coupled and uncoupled eigenstates of the Hamiltonian. The
expansion coefficients of uncoupled eigenstates into coupled
eigenstates (in an irreducible component of the tensor product), cfr.\
equation~(\ref{vv}), are again orthogonal polynomials (Jacobi,
continuous Hahn, or Hahn polynomials). 

All the relations given in section~II (deduced
in~\cite{Vanderjeugt,KV}) are obtained by Lie algebraic methods only. 
The purpose of the present paper is to examine the implications of
these relations once a realization for the Lie algebra $su(1,1)$ and
its representations ${\cal D}^+(k)$ are chosen. 
For $su(1,1)$, there exists a classical realization such that the
$J_0$-diagonal basis states $|k,n\rangle$ of ${\cal D}^+(k)$ are simply
the monomials $z^n$. This realization is considered in section~III. As
a consequence, the eigenstates of the Hamiltonian $H$ simply become
generating functions for the orthogonal polynomials (i.e.\ the
expansion coefficients), and also the $J_0$-diagonal basis states in
an irreducible component of the tensor product become a simple
functions. Then, it remains to consider the relation connecting coupled
and uncoupled eigenstates of the Hamiltonian in the tensor product, and
investigate its explicit form in this realization. For the case of
Laguerre and Jacobi polynomials, this is done in section~IV; the
outcome is a generalization of classical generating functions for
Laguerre polynomials. For the case of Meixner and Hahn polynomials,
this analysis is performed in section~V; the outcome is the Poisson
kernel formula for Meixner (or Meixner-Pollaczek) polynomials,
equation~(\ref{ser2}). Although the formula thus obtained is known, the
method used to reach it is original and interesting~: in this Lie
algebraic framework it is a simple consequence of a very general
expansion (equation~(\ref{vv})) in a particular realization.

Then we
turn our attention to the $q$-analog of the
results obtained so far. Since the method is purely algebraic, it works
not only for the Lie algebra $su(1,1)$ but also for the quantum algebra
$U_q(su(1,1))$~\cite{Burban}. In section~VI, the quantum algebra
$U_q(su(1,1))$ is defined, and the positive discrete series
representations with a standard basis are given. Next, a self-adjoint
operator playing the role of Hamiltonian is considered, and its
eigenstates are expanded in the standard basis with expansion
coefficients proportional to Al-Salam--Chihara polynomials. Just as for
the $su(1,1)$ case, coupled and uncoupled eigenstates of this operator
in the tensor product are constructed and the expansion of one into the
other is given in terms of Askey-Wilson polynomials. The results
presented in section~VI are mainly taken from~\cite{KV}. Just as for
$su(1,1)$, the purpose is now to investigate the implication of these
formulas once a realization is chosen. For $U_q(su(1,1))$, this
realization is given in section~VII, and the standard basis for
positive discrete series representations consist again of monomials $z^n$.  
Finally, the relation connecting coupled and uncoupled eigenstates is
shown to give rise to the (symmetric) Poisson kernel formula for
Al-Salam--Chihara polynomials in this realization. Although this
formula has recently been found by means of classical methods, the
present deduction is simple, purely algebraic, and a natural $q$-analog
of the Poisson kernel for Meixner-Pollaczek polynomials in the
$su(1,1)$ framework. 

In a final section, two different realizations of $su(1,1)$ and its
representations are considered. These realizations have been used in
physical models. We show again how our general results of section~II
have interesting implications for special functions. In particular,
integrals over products of Laguerre and Hermite polynomials are
expressed in terms of a Meixner polynomial. 
This section is completed by some general comments and further outlook.

\renewcommand{\theequation}{\arabic{section}.{\arabic{equation}}}
\setcounter{equation}{0}
\section{The Lie algebra $su(1,1)$ and positive discrete series
representations} 

The Lie algebra $su(1,1)$ is generated by $J_0, J_\pm$ subject to
the relations
\beq
[J_0, J_\pm]=\pm J_\pm,\qquad [J_+,J_-]=-2 J_0,
\label{defsu11}
\eeq
with the conditions $J_0^\dagger=J_0$ and $J_\pm^\dagger=J_\mp$. The
positive discrete representations~\cite{VK} ${\cal D}^+(k)$ are labeled
by a positive real number $k$. The representation space is
$\ell^2(\Zah_+)$, with orthonormal 
basis vectors denoted by $e^{(k)}_n$, with $n=0,1,2,\cdots$ (and
sometimes denoted by $k,n\rangle$).
The explicit action of the generators in this representation $(k)$ is
given by~:
\beq
 \begin{array}{l}
 J_0 e^{(k)}_n = (n+k) e^{(k)}_n ,\\
 J_+ e^{(k)}_n = \sqrt{(n+1)(2k+n)} e^{(k)}_{n+1}\\
 J_- e^{(k)}_n = \sqrt{n(2k+n-1)} e^{(k)}_{n-1}.
 \end{array}
\label{su11action}
\eeq

In two previous papers~\cite{Vanderjeugt,KV}, 
a recurrence operator in the Lie algebra
$su(1,1)$ was related to a Jacobi matrix. Rather than working with the
spectral theory of Jacobi matrices, it will be more appropriate for
this paper to use the (equivalent) notion of formal or generalized
eigenvectors, to be identified with eigenstates of a Hamiltonian. 
In~\cite{KV}, the formal eigenvectors
\beq
v^{(k)}(x)=\sum_{n=0}^\infty l^{(k)}_n(x) e^{(k)}_n
\label{formalv}
\eeq
of the self-adjoint operator $X=\sigma J_0-J_+-J_-$ ($\sigma\in\Real$) 
in the representation $(k)$ were studied; in a boson model such as
referred to in the introduction the Hamiltonian $H$ is essentially
equal to this operator $X$. The formal vector
$v^{(k)}(x)$ is a generalized eigenvector of $X$ for the eigenvalue
$\la(x)$ provided $l^{(k)}_n(x)$
satisfies a three-term recurrence relation~\cite{Vanderjeugt,KV}. 
This leads to an
identification of $l^{(k)}_n(x)$ with orthogonal polynomials in $x$.
These polynomials associated with $X$ are of different type for
$|\sigma|=2$, $|\sigma|<2$ or $|\sigma|>2$. Labelling the operator in
the three distinct cases as follows~:
\bea
X_2&=& 2J_0-J_+-J_- ,\label{X2}\\
X_\phi&=&-2\cos\phi\;J_0+J_++J_-,\qquad 0<\phi<\pi,\label{Xphi}\\
X_c&=&-(c+1/c)J_0+J_++J_-,\qquad 0<c<1, \label{Xc}
\eea
we can give a summary of some results obtained earlier in Table~1. The
spectrum of $X$ is determined by the support of the measure of the
associated orthogonal polynomials, being continuous for $|\sigma|=2$
and $|\sigma|<2$, and discrete for $|\sigma|>2$. 

\begin{table}[htb]
\caption{Orthogonal polynomials appearing in the formal eigenvector $v^{(k)}(x)$.} 
\label{tab1}
\[
\begin{tabular}{|c|c|c|c|c|}
\hline
$X$ & $l^{(k)}_n(x)$ & polynomial & eigenvalue $\la(x)$ & spectrum\\ \hline
& & & & \\[-2mm]
$X_2$ & $\sqrt{n!\over(2k)_n}L^{(2k-1)}_n(x)$ &
generalized Laguerre & $x$ & $0\leq x<\infty$ \\[3mm]
$X_\phi$ &$\sqrt{n!\over\Gamma(2k+n)}P^{(k)}_n(x;\phi)$ & Meixner-Pollaczek &
$2x\sin\phi$ & $x\in\Real$ \\[3mm]
$X_c$ & $\sqrt{(2k)_n\over n!}c^n M_n(x;2k;c^2)$ &
Meixner & $(c-1/c)(k+x)$ & $x\in\Nat$ \\[3mm] \hline
\end{tabular}
\]
\end{table}

In Table~1, $(\alpha)_n=\Gamma(\alpha+n)/\Gamma(\alpha)$ is the common
notation for the Pochhammer symbol in terms of the classical $\Gamma$
function. The definition of the orthogonal polynomials in terms of
hypergeometric series is as follows~\cite{KS}~:
\bea
L^{(\al)}_n(x) & = & {(\al+1)_n\over n!} {\;}_1F_1\left[{-n\atop
\al+1}; x\right],\qquad (\al>-1);\label{Laguerre}\\
P^{(\la)}_n(x;\phi)&=&{(2\la)_n\over n!}e^{i
n\phi}{\;}_2F_1\left[{-n,\la+ix \atop 2\la};1-e^{-2i\phi}\right],
\qquad (\la>0);\label{MP}\\
M_n(x;\be;c)&=& {\;}_2F_1\left[{-n,-x \atop \be};1-1/c\right],
\qquad (\be>0).\label{Meixner}
\eea
The notation for hypergeometric series is the standard
one~\cite{Slater,GR}.

By considering the tensor product decomposition for $su(1,1)$
representations, new summation formulas or convolution theorems were
obtained for these polynomials. The tensor product decomposes as
follows~\cite{VK}~: 
\beq
(k_1)\otimes(k_2) = \bigoplus_{j=0}^\infty (k_1+k_2+j).
\label{tensdec}
\eeq
The ``coupled basis vectors'' are written in terms of the uncoupled
ones by means of the Clebsch-Gordan coefficients~:
\beq
e^{(k_1k_2)k}_n = \sum_{n_1,n_2} C^{k_1,k_2,k}_{n_1,n_2,n}\ 
e^{(k_1)}_{n_1} \otimes e^{(k_2)}_{n_2}.
\label{CGC}
\eeq
Herein, $k=k_1+k_2+j$ for some integer $j\geq 0$, and the sum is such
that $n_1+n_2=j+n$. 
Explicit expression for the Clebsch-Gordan coefficients are given,
e.g.\ in Ref.~\cite{Vanderjeugt}. 

In the tensor product space, the generalized eigenvectors of
$\Delta(X)=X\otimes 1+1\otimes X$ can be considered. Again, there are
``uncoupled'' eigenvectors $v^{(k_1)}(x_1) v^{(k_2)}(x_2)$ with
eigenvalue $\la(x_1+x_2)$, but also ``coupled'' eigenvectors
$v^{(k_1k_2)k}(x_1+x_2)$ defined as follows~:
\beq
v^{(k_1k_2)k}(x_1+x_2)=\sum_{n=0}^\infty l^{(k)}_n(x_1+x_2)
e^{(k_1k_2)k}_n,
\label{vcoup}
\eeq
where $k=k_1+k_2+j$ for some integer $j\geq0$. It was shown
in~\cite{Vanderjeugt,KV} 
that
\beq
v^{(k_1)}(x_1) v^{(k_2)}(x_2) = \sum_{j=0}^\infty S_j^{(k_1,k_2)}(x_1,x_2)
v^{(k_1k_2)k_1+k_2+j}(x_1+x_2),
\label{vv}
\eeq
where $S$ is again an orthogonal polynomial given in Table~2, according
to which of the three cases is considered. The constants in this table
are determined by
\bea
C_1&=&\left(j!/((2k_1)_j(2k_2)_j(2k_1+2k_2+j-1)_j)\right)^{1/2},\label{C1}\\ 
C_2&=&\left(j!(2k_1+2k_2+2j-1)\Gamma(2k_1+2k_2+j-1)/
(\Gamma(2k_1+j)\Gamma(2k_2+j) )\right)^{1/2},\label{C2}\\
C_3&=&C_2 (2k_1)_j/j!.\label{C3}
\eea

\begin{table}[htb]
\caption{Orthogonal polynomial appearing in the expansion (2.13).}
\label{tab2}
\[
\begin{tabular}{|c|c|c|}
\hline
$X$ & $S_j^{(k_1,k_2)}(x_1,x_2)$ & polynomial \\ \hline
 & & \\[-2mm]
$X_2$ & $C_1 (-1)^j (x_1+x_2)^j P_j^{(2k_1-1,2k_2-1)}\left({x_2-x_1\over
x_2+x_1} \right)$ & Jacobi \\[3mm] 
$X_\phi$ & $C_2 (-2\sin\phi)^j
p_j(x_1;k_1,k_2-i(x_1+x_2),k_1,k_2+i(x_1+x_2))$ & 
continuous Hahn \\[3mm] 
$X_c$ & $C_3(-c+1/c)^j (-x_1-x_2)_j \; Q_j(x_1;2k_1-1,2k_2-1,x_1+x_2)$ &
Hahn\\[3mm] \hline
\end{tabular}
\]
\end{table}

The polynomials appearing in Table~2 are defined as follows~\cite{KS}~:
\bea
P^{(\al,\be)}_n(x) &=& {(\al+1)_n\over
n!}{\;}_2F_1\left[{-n,n+\al+\be+1 \atop \al+1};{1-x\over 2}\right],
\label{Jacobi} \\
p_n(x;a,b,c,d)&=&i^n{(a+c)_n(a+d)_n\over
n!}{\;}_3F_2\left[{-n,n+a+b+c+d-1,a+ix \atop
a+c,a+d};1\right],\label{cHahn} \\
Q_n(x;a,b,N)&=&{\;}_3F_2\left[{-n,n+a+b+1,-x \atop a+1,-N};1\right].
\label{Hahn} 
\eea
A number of interesting new convolution identities for orthogonal
polynomials were constructed in~\cite{KV} from the relation ($k=k_1+k_2+j$)
\beq
\sum_{n_1+n_2=n+j} C^{k_1,k_2,k}_{n_1,n_2,n}\  l^{(k_1)}_{n_1}(x_1) 
l^{(k_2)}_{n_2}(x_2) = l^{(k)}_{n}(x_1+x_2) S_j^{(k_1,k_2)}(x_1,x_2).
\label{ident}
\eeq
In the simplest case, where $l^{(k)}_n(x)$ corresponds to a generalized
Laguerre polynomial and $S_j^{(k_1,k_2)}(x_1,x_2)$ to a Jacobi polynomial,
this relation reduces to~\cite[(1.1)]{Vanderjeugt}. Recently, this
equation has been applied to find transformation brackets for $U(N)$
boson models~\cite{Kota}.

\setcounter{equation}{0}
\section{Realization of $su(1,1)$}

In this section we give a realization of $su(1,1)$ and of the discrete
series representation ${\cal D}^+(k)$. Assume that $k>1/2$, and
consider the Hilbert space of analytic functions $f(z)$ ($z\in\C$) on
the unit disc $|z|<1$, with inner product~\cite{VK}
\beq
(f_1,f_2)={2k-1\over\pi}\int\int_{|z|<1} f_1(z)\overline{f_2(z)}
(1-|z|^2)^{2k-2} dxdy,\qquad (z=x+iy).
\label{inprod}
\eeq
The following functions form an orthonormal basis~:
\beq
e^{(k)}_n \equiv e^{(k)}_n(z) = \sqrt{(2k)_n\over n!} z^n,
\label{e-real}
\eeq
since $(e^{(k)}_n,e^{(k)}_m)=\delta_{m,n}$. The realization of the
$su(1,1)$ basis elements reads as follows~:
\beq
J_0=z{d\over dz}+k,\qquad J_-={d\over dz},\qquad J_+=z^2{d\over
dz}+2kz. 
\label{j-real}
\eeq
It is easy to verify that the action of these operators on the basis
(\ref{e-real}) is indeed the same as in (\ref{su11action}).

Next, consider the formal vectors~(\ref{formalv}), which are now
denoted by 
$v^{(k)}(x,z)$ since in this realization the basis vectors become
functions of $z$. Since the vectors $e^{(k)}_n$ are proportional to
$z^n$, one can see that the $v^{(k)}(x,z)$ are in fact generating
functions for the polynomials $l^{(k)}_n(x)$. The explicit forms of the
vectors is given in Table~3, for each of the three cases consider here.

\begin{table}[htb]
\caption{Explicit form of the vectors $v^{(k)}(x,z)$ in the present
realization.}
\label{tab3}
\[
\begin{tabular}{|c|c|c|}
\hline
$X$ & series  & generating function\\ \hline
 & & \\[-2mm]
$X_2$ & $\sum_{n=0}^\infty L_n^{(2k-1)}(x)z^n$ &
$(1-z)^{-2k}e^{xz/(z-1)}$ \\[3mm] 
$X_\phi$ & $\Gamma(2k)^{-1}\sum_{n=0}^\infty P_n^{(k)}(x;\phi)z^n$ & 
$(1-e^{i\phi}z)^{-k+ix}(1-e^{-i\phi}z)^{-k-ix}/\Gamma(2k)$ \\[3mm] 
$X_c$ & $\sum_{n=0}^\infty {(2k)_n c^n\over n!}M_n(x;2k,c^2)z^n$ &
$(1-z/c)^x(1-cz)^{-x-2k}$\\[3mm] \hline
\end{tabular}
\]
\end{table}

For each of the cases, the generating function can be found in
Ref.~\cite{KS}. One can also find it by explicitly solving the first
order differential equation 
\beq
X v^{(k)}(x,z) = \la(x) v^{(k)}(x,z),
\label{X-action}
\eeq
where $X$ is one of (\ref{X2})-(\ref{Xc}), in the realization
(\ref{j-real}).  

Next, consider the tensor product of two representations $(k_1)\otimes
(k_2)$ in the present realization, and the coupled vector
\beq
e^{(k_1k_2)k}_n(z_1,z_2) = \sum_{n_1,n_2} C^{k_1,k_2,k}_{n_1,n_2,n}\ 
e^{(k_1)}_{n_1}(z_1) \; e^{(k_2)}_{n_2}(z_2).
\label{CGC-real}
\eeq
Using properties of the generating function for the Clebsch-Gordan
coefficients, this expression can be written as follows~:
\bea
&&e^{(k_1k_2)k}_n(z_1,z_2)= \left[ {(2k_1)_j(2k_2)_j(2k_1+2k_2+2j)_n\over
j!n!(2k_1+2k_2+j-1)_j }\right]^{1/2} \nn\\
&&\times (-z_1+z_2)^j z_1^n\; {}_2F_1\left[{-n,2k_2+j \atop
2k_1+2k_2+2j}; 1-z_2/z_1\right],
\label{coup-real}
\eea
where, as before, $k=k_1+k_2+j$.

In the following sections we shall consider the explicit forms of
(\ref{vcoup}) and (\ref{vv}) in this realization, and its consequences.

\setcounter{equation}{0}
\section{The case of Jacobi polynomials}

Consider equation (\ref{vv}) in the case $X=X_2$,
\beq
v^{(k_1)}(x_1,z_1) v^{(k_2)}(x_2,z_2) = \sum_{j=0}^\infty 
S_j^{(k_1,k_2)}(x_1,x_2)
v^{(k_1k_2)k_1+k_2+j}(x_1+x_2,z_1,z_2);
\label{vv-real}
\eeq
in this situation the polynomials $S_j$ are Jacobi polynomials. Using
the formulas of Table~2, and the new variables
\beq
\begin{array}{l}
s=x_2+x_1,\\
r=(x_2-x_1)/(x_2+x_1);
\end{array} \qquad\qquad
\begin{array}{l}
x_1=s(1-r)/2,\\
x_2=s(1+r)/2;
\end{array} 
\label{s-r}
\eeq
this equation becomes
\beq
v^{(k_1)}(s(1-r)/2,z_1) v^{(k_2)}(s(1+r)/2,z_2) = \sum_{j=0}^\infty 
C_1 (-1)^j s^j P_j^{(2k_1-1,2k_2-1)}(r) 
v^{(k_1k_2)k_1+k_2+j}(s,z_1,z_2).
\label{tmp1}
\eeq
Now we can integrate with respect to the weight function of Jacobi
polynomials~: 
\bea
&&\int_{-1}^{1} P_j^{(2k_1-1,2k_2-1)}(r) v^{(k_1)}({s(1-r)\over 2},z_1)
v^{(k_2)}({s(1+r)\over 2},z_2) (1-r)^{2k_1-1}(1+r)^{2k_2-1}dr \nn\\
&&\qquad = C_1 h_j (-1)^j s^j \; v^{(k_1k_2)k_1+k_2+j}(s,z_1,z_2),
\label{int-Jacobi}
\eea
where $h_j$ is the norm of the Jacobi polynomials of degree $j$, i.e.
\beq
h_j={2^{2k_1+2k_2-1}\over 2k_1+2k_2+2j-1} {\Gamma(2k_1+j)\Gamma(2k_2+j)
\over j!\Gamma(2k_1+2k_2+j-1)}.
\label{Jacobi-norm}
\eeq
Using the expressions for the functions $v$ given in Table~3, we obtain
\bea
&& C_1 h_j (-1)^j s^j \; v^{(k_1k_2)k_1+k_2+j}(s,z_1,z_2) =
(1-z_1)^{-2k_1} (1-z_2)^{-2k_2}e^{{s\over 2}({z_1\over z_1-1}+
{z_2\over z_2-1})} \nn\\
&\times&\int_{-1}^{1} P_j^{(2k_1-1,2k_2-1)}(r) e^{s(z_1-z_2)\over
2(z_1-1)(z_2-1)} (1-r)^{2k_1-1}(1+r)^{2k_2-1}dr.
\label{tmp2}
\eea

\begin{lemm}
\bea
&&I=\int_{-1}^{1}(1-r)^a (1+r)^b P_j^{(a,b)}(r) e^{cr}dr = \nn\\
&&2^{a+b+1}{\Gamma(a+j+1)\Gamma(b+j+1)\over j!\Gamma(a+b+2j+2)} e^{-c}
(2c)^j \; {}_1F_1\left[{b+j+1\atop a+b+2j+2};2c\right].
\label{integral1}
\eea
\end{lemm}

\noindent {\em Proof.} Writing
\beq
P_j^{(a,b)}(r) = {1\over 2^j} \sum_{m=0}^j \left(j+a\atop m\right)
\left(j+b \atop j-m\right) (r-1)^{j-m}(r+1)^m,
\eeq
and using
\bea
&&\int_{-1}^{1} (1-r)^\al(1+r)^\be e^{\ga r}dr = \nn\\
&& 2^{\al+\be+1} {\Gamma(\al+1)\Gamma(\be+1)\over\Gamma(\al+\be+2)}
e^{-\ga} {}_1F_1\left[{\be+1\atop \al+\be+2};2\ga\right],
\eea
one obtains, after some simplifications,
\beq
I=(-1)^j 2^{a+b+1}e^{-c}{\Gamma(a+j+1)\Gamma(b+j+1)\over
j!\Gamma(a+b+j+2)} \sum_{j=0}^m {(-j)_m\over m!} {}_1F_1\left[{b+m+1
\atop a+b+j+2};2c\right].
\label{tmp3}
\eeq
In the sum over $m$, the ${}_1F_1$ is written explicitly as a series,
the two summation variables are changed, and then Vandermonde's
summation theorem can be applied; putting $\al=a+b+j+2$ and $\be=b+1$,
this gives explicitly~:
\bea
&&\sum_{j=0}^m {(-j)_m\over m!} {}_1F_1\left[{\be+m
\atop \al};\ga\right]=\sum_{m=0}^j {(-j)_m\over m!} \sum_{k=0}^\infty
{(\be+m)_k \ga^k \over (\al)_k k!} \nn\\
&&=\sum_{k=0}^\infty {(\be)_k \ga^k \over (\al)_k k!} \sum_{m=0}^j
{(b+k)_m (-j)_m \over (b)_m m!} \nn\\
&&=\sum_{k=0}^\infty {(\be)_k \ga^k \over (\al)_k k!} 
{(-k)_j \over (b)_j }.
\eea
In the last sum $k$ goes from $j$ upto $\infty$, so let $l=k-j$, and
rewrite the series; one finds
\beq
{(-1)^j \ga^j \over(\al)_j}\sum_{l=0}^\infty {(\be+j)_l \ga^l \over
(\al+j)_l l!} = {(-1)^j \ga^j \over(\al)_j} {}_1F_1\left[{\be+j \atop
\al+j};\ga \right].
\eeq
Putting this back in (\ref{tmp3}) finally proves the lemma.

As a consequence, we now have the following explicit form for
$v^{(k_1k_2)k_1+k_2+j} (s,z_1,z_2)$~:
\bea
&&v^{(k_1k_2)k_1+k_2+j} (s,z_1,z_2)=(-1)^j \left[{(2k_1)_j(2k_2)_j\over
j!(2k_1+2k_2+j-1)_j} \right]^{1/2} (1-z_1)^{-2k_1-j}(1-z_2)^{-2k_2-j}
\nn\\ 
&&\times (z_1-z_2)^j e^{sz_1/(z_1-1)}{\;}_1F_1\left[ {2k_1+j \atop
2k_1+2k_2+2j} ; {s(z_1-z_2)\over(z_1-1)(z_2-1)}\right].
\label{v1}
\eea
On the other hand, we know from (\ref{vcoup}) that ($k=k_1+k_2+j$)~:
\beq
v^{(k_1k_2)k}(s,z_1,z_2)=\sum_{n=0}^\infty l^{(k)}_n(s)
e^{(k_1k_2)k}_n(z_1,z_2),
\label{v2}
\eeq
with $l^{(k)}_n(s)$ given in Table~1 and $e^{(k_1k_2)k}_n(z_1,z_2)$
given in (\ref{coup-real}). Thus we obtain,
\bea
&&\sum_{n=0}^\infty L_n^{(2k)}(s)z_1^n{\;}_2F_1\left[{-n,2k_2+j\atop
2k_1+2k_2+2j}; 1-z_2/z_1\right]= \nn\\
&&(1-z_1)^{-2k_1-j}(1-z_2)^{-2k_2-j}
e^{sz_1/(z_1-1)}{\;}_1F_1\left[ {2k_1+j \atop
2k_1+2k_2+2j} ; {s(z_1-z_2)\over(z_1-1)(z_2-1)}\right].
\label{series-Laguerre}
\eea
This can be rewritten in a more appropriate form~:

\begin{prop}
\bea
&&\sum_{n=0}^\infty L_n^{(b-1)}(s)z_1^n{\;}_2F_1\left[{-n,a\atop
b}; 1-{z_2\over z_1}\right]=  \nn\\
&& \quad (1-z_1)^{a-b}(1-z_2)^{-a}
e^{sz_1/(z_1-1)}{\;}_1F_1\left[ {a \atop
b} ; {s(z_1-z_2)\over(z_1-1)(z_2-1)}\right].
\label{series2-Laguerre}
\eea
\end{prop}

This is a generalization of two classical generating functions for the
Laguerre polynomials. For $a=0$, or $z_1=z_2$,
it reduces to the first classical result 
\beq
\sum_{n=0}^\infty L_n^{(b-1)}(s)z_1^n = (1-z_1)^{-b}e^{sz_1/(z_1-1)};
\eeq
for $z_2=0$, it reduces to
\beq
\sum_{n=0}^\infty L_n^{(b-1)}(s){(b-a)_n\over(b)_n} z_1^n = 
(1-z_1)^{-b}{\;}_1F_1\left[ {b-a \atop b} ; {sz_1\over(z_1-1)}\right].
\eeq

Note that (\ref{series2-Laguerre}) can be written in a more general form~:
\bea
&&\sum_{n=0}^\infty {\;}_1F_1\left[{-n\atop b};x\right]
{\;}_2F_1\left[{-n,a\atop b}; y\right]{(b)_n\over n!}z^n = \nn\\
&&(1-z)^{a-b}(1-z+yz)^{-a}
e^{xz/(z-1)}{\;}_1F_1\left[ {a \atop
b} ; {xyz\over(1-z)(1-z+yz)}\right].
\label{ser1}
\eea

If one is not interested in the particular form of
$v^{(k_1k_2)k_1+k_2+j} (s,z_1,z_2)$, but only in comparing the two
expressions of this vector in order to obtain (\ref{series-Laguerre}),
many simplifications take place. In particular, note that
(\ref{series-Laguerre}) does not explicitly depend upon $j$ but only
upon $2k_1+j$ and $2k_2+j$. Thus, one can obtain the same formula
(\ref{series2-Laguerre}) just by working out only the vector with
$j=0$. This will be done for the remaining cases in the following
sections.

\setcounter{equation}{0}
\section{The case of Meixner and Meixner-Pollaczek polynomials}

Consider the case $X=X_c$; then (\ref{vv-real}) holds with $S_j$ proportional
to a Hahn polynomial. First, assume that $s=x_1+x_2$ is a positive
integer (later this condition will disappear by analytic continuation).
We can then apply the (discrete) orthogonality
relation for Hahn polynomials, and obtain the $j=0$ component
\beq
v^{(k_1k_2)k_1+k_2}(s,z_1,z_2) {(2k+1+2k_2-1)_{s+1}\over
s!(2k_1+2k_2-1)} = \sum_{x_1=0}^s {(2k_1)_{x_1} (2k_2)_{s-x_1} \over
x_1!(s-x_1)!} v^{(k_1)}(x_1,z_1) v^{(k_2)}(s-x_1,z_2).
\label{m-tmp1}
\eeq
With the explicit forms for $v^{(k_1)}$ and $v^{(k_2)}$ from Table~3,
this gives rise to
\beq
v^{(k_1k_2)k_1+k_2}(s,z_1,z_2)=(1-cz_1)^{-2k_1}(1-z_2/c)^s(1-cz_2)^{-s-2k_2}
{\;}_2F_1\left[ {-s,2k_1\atop 2k_1+2k_2};1-{z_2\over z_1}\right].
\label{m-v1}
\eeq
On the other hand, (\ref{v2}) now becomes
\beq
v^{(k_1k_2)k_1+k_2}(s,z_1,z_2)=\sum_{n=0}^\infty {(2k_1+2k_2)_n\over n!}
M_n(s;2k_1+2k_2,c^2) c^n z_1^n {\;}_2F_1\left[ {-n,2k_2\atop
2k_1+2k_2};1-{z_2\over z_1}\right] .
\label{m-v2}
\eeq
Applying a series transformation on the last $_2F_1$  yields
\bea
&&\sum_{n=0}^\infty {(b)_n\over n!}
M_n(s;b,c^2) c^n z_2^n {\;}_2F_1\left[ {-n,a\atop
b};1-{z_2\over z_1}\right] = \nn\\
&&(1-cz_1)^{-a}(1-cz_2)^{a-b}\left({1-z_2/c\over 1-cz_2}\right)^s
{\;}_2F_1\left[ {-s,a\atop
b};{(c-1/c)(z_2-z_1)\over(1-cz_1)(1-z_2/c)}\right] .
\label{series-Meixner}
\eea
Given the Meixner polynomials in terms of a $_2F_1$, this leads to
the following series expression~:

\begin{prop}
\bea
&&\sum_{n=0}^\infty 
{\;}_2F_1\left[ {-n,a\atop c};x\right] 
{\;}_2F_1\left[ {-n,b\atop c};y\right] {(c)_n z^n\over n!}
= \nn\\
&&(1-z+xz)^{-a}(1-z+yz)^{-b}(1-z)^{a+b-c}
{\;}_2F_1\left[ {a,b\atop
c};{xyz\over(1-z+xz)(1-z+yz)}\right] .
\label{ser2}
\eea
\end{prop}

This formula was already given in Ref.~\cite{Er}, p.~85, eq.~(12), and
can be interpreted as the Poisson kernel for Meixner or
Meixner-Pollaczek polynomials~\cite{KV2}. Observe that (\ref{ser1}) is
a limiting case of (\ref{ser2})~: putting $x=x'/a$ in (\ref{ser2}) and
taking the limit $a\rightarrow\infty$ leads to (\ref{ser1}).
 
The case of Meixner-Pollaczek polynomials leads essentially to the same
formulas.

\setcounter{equation}{0}
\section{The algebra $U_q(su(1,1))$}

Let $U_{q}(sl(2,\C))$ ($0<q<1$) be the complex unital
associative algebra generated by $A$, $B$, $C$, $D$ subject to the
relations\footnote{Compared to Ref.~\cite{KV}, $q$ is replaced by
$q^{1/2}$ in order to have $q$ as basis of the basic hypergeometric
series appearing later.}
\beq
AD=1=DA, \quad AB=q^{1/2}BA,\quad AC=q^{-1/2}CA,\quad
BC-CB = {{A^2-D^2}\over{q^{1/2}-q^{-1/2}}}.
\label{defrel}
\eeq
This algebra can be equiped with a comultiplication, counit, and
antipode, turning it into a Hopf algebra~\cite{CP,Burban}. 
The Hopf $*$-algebra 
$U_q(su(1,1))$ has the following $*$-structure~:
\beq
A^*=A,\quad B^*=-C,\quad C^*=-B, \quad D^*=D.
\label{star}
\eeq
The positive discrete representations are labeled by a positive real number
$k$. The representation space is again $\ell^2(\Zah_+)$, with orthonormal
basis vectors denoted by $e^{(k)}_n$, with $n=0,1,2,\cdots$.
The explicit action of the generators in this representation $(k)$ is
given by~\cite{KV}
\bea
&&A\, e^{(k)}_n =q^{(k+n)/2}\, e^{(k)}_n, \label{A}\\
&&C\, e^{(k)}_n = q^{(1-2k-2n)/4}
{{\sqrt{(1-q^{n})(1-q^{2k+n-1})}}\over{q^{1/2}-q^{-1/2}}}\,
e^{(k)}_{n-1}, \label{C}\\
&&B \, e^{(k)}_n = q^{-(1+2k+2n)/4}
{{\sqrt{(1-q^{n+1})(1-q^{2k+n})}}\over{q^{-1/2}-q{1/2}}}\,
e^{(k)}_{n+1}. \label{B} 
\eea
Let $s\in\Real\backslash\{0\}$, and define
\beq
Y_s = q^{1/4}B-q^{-1/4}C + {{s^{-1}+s}\over{q^{-1/2}-q^{1/2}}}(A-D).
\label{Ys}
\eeq
Under the comultiplication
\def\De{\Delta}
\bea
\De(A)=A\otimes A,\quad \De(B)=A\otimes B+B\otimes D,\nn \\
\De(C) = A\otimes C+C\otimes D, \quad \De(D)=D\otimes D,
\label{comult}
\eea
$Y_s$ is twisted primitive, and $Y_sA$ is a self-adjoint element. 
Here, $Y_sA$ is playing the role of the operator $X$ for $su(1,1)$.
In a previous paper~\cite{KV}, the formal eigenvectors 
\beq
v^{(k)}(x)=\sum_{n=0}^\infty l^{(k)}_n(x) e^{(k)}_n
\label{q-fv}
\eeq
of $Y_sA$ have been obtained. In order to express the results
of~\cite{KV}, recall the definition of Askey-Wilson~\cite{AW} 
and Al-Salam and Chihara polynomials~\cite{AC}. 
The Askey-Wilson polynomial is
defined by ($x=\cos\th$)~:
\beq
p_m(x;a,b,c,d|q) = a^{-m} (ab,ac,ad;q)_m\, {}_4\vp_3
\left[ {{q^{-m},abcdq^{m-1},ae^{i\th},ae^{-i\th}}\atop
{ab,\ ac,\ ad}}; q,q\right]
\label{a-w}
\eeq
and it is symmetric in its parameters $a$, $b$, $c$ and $d$.
In (\ref{a-w}), the notation for basic hypergeometric series and for
the shifted $q$-factorial is taken from~\cite{GR}.
The Al-Salam and Chihara polynomials are obtained by taking
$c=d=0$ in the Askey-Wilson polynomials;
\beq
s_m(x;a,b|q) = p_m(x;a,b,0,0|q) = a^{-m} (ab;q)_m\,
{}_3\vp_2 \left[ {{q^{-m},ae^{i\th},ae^{-i\th}}\atop{ab,\ 0}};
q,q\right].
\label{a-c}
\eeq
Using the abbreviation $\mu(x)=(x+x^{-1})/2$, it was shown in \cite{KV}
that for 
\beq
l^{(k)}_n(x) = {1\over \sqrt{(q,q^{2k};q)_n}} s_n(\mu(x);q^ks,q^k/s|q),
\label{l-a-c}
\eeq
the vectors (\ref{q-fv}) are formal eigenvectors of $Y_sA$ for the
eigenvalue 
$$
\la(x)=2(\mu(s)-\mu(x))/(q^{1/2}-q^{-1/2}).
$$ 

The tensor product of two $U_q(su(1,1))$ representations
$(k_1)\otimes(k_2)$ is the same as in (\ref{tensdec}), and the Clebsch-Gordan
coefficients in
\beq
e^{(k_1k_2)k}_n = \sum_{n_1,n_2} C^{k_1,k_2,k}_{n_1,n_2,n}\ 
e^{(k_1)}_{n_1} \otimes e^{(k_2)}_{n_2},
\label{q-CGC}
\eeq
have been determined, e.g., in Ref.~\cite{KV}.  In the tensor product
space, the generalized eigenvectors of $\De(Y_sA)$ have been
considered, both in coupled and uncoupled form. It was shown that
\beq
v^{k_1;k_2}(x_1,x_2)=\sum_{n_1,n_2}
{s_{n_1}(\mu(x_1);q^{k_1}x_2,q^{k_1}/x_2|q)\over\sqrt{
(q,q^{2k_1};q)_{n_1}}} 
{s_{n_2}(\mu(x_2);q^{k_2}s,q^{k_2}/s|q)\over\sqrt{
(q,q^{2k_2};q)_{n_2}}} e^{(k_1)}_{n_1} \otimes e^{(k_2)}_{n_2}
\label{q-v1}
\eeq
is a generalized eigenvector of $\De(Y_sA)$ for the eigenvalue
$\la(x_1)$. In ``coupled'' form, 
\beq
v^{(k_1k_2)k}(x_1)=\sum_{n=0}^\infty
{s_{n}(\mu(x_1);q^{k}s,q^{k}/s|q)\over\sqrt{
(q,q^{2k};q)_{n}}} e^{(k_1k_2)k}_n,
\label{q-v2}
\eeq
is a generalized eigenvector of $\De(Y_sA)$ for the same eigenvalue. The
relation between the two generalized eigenvectors was given
in~\cite{KV}~: 
\beq
v^{k_1;k_2}(x_1,x_2)=\sum_{j=0}^\infty C_j p_j(\mu(x_2);q^{k_1}x_1,
q^{k_1}/x_1, q^{k_2}s, q^{k_2}/s|q) v^{(k_1k_2)k_1+k_2+j}(x_1),
\label{q-vv}
\eeq
where $p_j$ is an Askey-Wilson polynomial and 
\beq
C_j=\left( (q,q^{2k_1},q^{2k_2},q^{2k_1+2k_2+j-1};q)_j \right)^{-1/2}.
\label{Cj}
\eeq

\setcounter{equation}{0}
\section{Realization of $U_q(su(1,1))$}

A realization of $U_q(su(1,1))$ and the representation labelled by
$(k)$ is closely related to the one given for $su(1,1)$ in Section~III.
The Hilbert space is the same (up to a rescaling of the inner product),
and the basis functions are now
\beq
e^{(k)}_n\equiv e^{(k)}_n(z)=\sqrt{ (q^{2k};q)_n\over(q;q)_n} z^n.
\label{qe-real}
\eeq
The realization of the $U_q(su(1,1))$ generators is given in terms of
the operators
\beq
T_q f(z)=f(qz),\qquad D_q ={1-T_q\over(1-q)z},
\label{oper}
\eeq
and reads
\bea
&&A= q^{k/2} T_{q^{1/2}},\quad D= q^{-k/2} T_{q^{-1/2}},\label{qA}\\
&&B= q^{(1-2k)/4}\left( z^2 D_q T_{q^{-1/2}} + {1-q^{2k} \over 1-q} z
T_{q^{1/2}} \right),\label{qB}\\
&&C=-q^{(3-2k)/4} D_q T_{q^{-1/2}}. \label{qC}
\eea

For this realization, the formal vectors (\ref{q-fv}) become
\bea
v^{(k)}(x,z)&=&\sum_{n=0}^\infty s_n(\mu(x);q^ks,q^k/s|q)
{z^n\over(q;q)_n}\\ 
&=& {(q^k zs, q^kz/s;q)_\infty \over (zx,z/x;q)_\infty},
\label{qv-real}
\eea
following from a known generating function for the Al-Salam and Chihara
polynomials~\cite{KS}. 
We also wish to obtain an explicit form for the coupled
vectors $e^{(k_1k_2)k}_n(z_1,z_2)$. Using the proof of Lemma~4.4
of~\cite{KV}, one finds~($k=k_1+k_2+j$)~:
\bea
&&e^{(k_1k_2)k}_n(z_1,z_2) = q^{-nj-nk_1} z_2^{n+j}
(q^{k_1}z_1/z_2;q)_j \sqrt{(q^{2k_1},q^{2k_2};q)_j \over
(q,q^{2k_1+2k_2+j-1};q)_j} \nn\\
&&\times \sqrt{(q^{2k_1+2k_2+2j};q)_n\over(q;q)_n} 
\ {}_3\vp_2 \left[ {{q^{-n},q^{2k_1+j}, q^{k_1+j}z_1/z_2}\atop{
q^{2k_1+2k_2+2j},\ 0}};q,q\right].
\label{q-coup}
\eea

\setcounter{equation}{0}
\section{Poisson kernel for Al-Salam--Chihara polynomials}

In the realization of the previous section, the formal vectors defined
in (\ref{q-v1}) can be rewritten using (\ref{qv-real}) and they become
\beq
v^{k_1;k_2}(x_1,x_2,z_1,z_2) = {(q^{k_1}z_1 x_2,q^{k_1}z_1/ x_2,
q^{k_2}z_2 s,q^{k_2}z_2/ s; q)_\infty \over (z_1 x_1,z_1/x_1,z_2 x_2,
z_2/x_2;q)_\infty}.
\label{qv1-real}
\eeq
Note that in all the previous formulas the variables of the Al-Salam
and Chihara (or Askey-Wilson) polynomials are $\mu(x_1)$ or $\mu(x_2)$;
we can represent $\mu(x_l)=\cos\th_l$ and thus $x_l=e^{i\th_l}$ for
some real $\th_l$ ($l=1,2$). The purpose is now to integrate equation
(\ref{q-vv}) so as to obtain an explicit form for
$v^{(k_1k_2)k_1+k_2+j}(x_1,z_1,z_2)$ for $j=0$. Let
\beq
(a,b,c,d)=(q^{k_1}x_1, q^{k_1}/x_1,q^{k_2}s, q^{k_2}/s).
\label{abcd}
\eeq
If we assume that $q^{k_2}<|s|<q^{-k_2}$, then $\max\{|a|,|b|,|c|,|d|\}
<1$, and the Askey-Wilson polynomial in (\ref{q-vv}) has an absolute
continuous measure. Using the notation
\beq
h(x;a)=h(x;a;q)=(ae^{i\th},ae^{-i\th};q)_\infty,\qquad\hbox{for}\
x=\cos\th,
\label{h1}
\eeq
and
\beq
h(x;a,b,c,d)=h(x;a)h(x;b)h(x;c)h(x;d),
\label{h2}
\eeq
the weight function for the Askey-Wilson polynomials reads~\cite[\S
7.5]{GR} 
\beq
w(x)\equiv w(x;a,b,c,d)={h(x;1,-1,q^{1/2},q^{-1/2})\over\sqrt{1-x^2}
h(x;a,b,c,d)} .
\label{w}
\eeq
The orthogonality relation for $p_n(x)\equiv p_n(x;a,b,c,d|q)$ is
\beq
\int_{-1}^1 p_m(x) p_n(x) w(x) dx = {\delta_{m,n}\over h_n},
\label{orth}
\eeq
where
\beq
h_n=h_0 {(abcdq^{-1};q)_n(1-abcdq^{2n-1}) \over (1-abcdq^{-1})(q,ab,ac,
ad,bc, bd,cd;q)_n},
\label{hn}
\eeq
and
\beq
h_0={(q,ab,ac,ad,bc,bd,cd;q)_\infty \over 2\pi(abcd;q)_\infty}.
\label{h0}
\eeq
It thus follows from (\ref{q-vv}) that 
\beq
v^{(k_1k_2)k_1+k_2}(x_1,z_1,z_2) = h_0 \int_{-1}^1 w(\mu(x_2);a,b,c,d)
v^{k_1;k_2}(x_1, x_2,z_1, z_2) d\mu(x_2).
\label{Int1}
\eeq
Using the explicit form (\ref{qv1-real}), the notation 
\beq
f=z_2, \quad g=q^{k_1} z_1,
\label{fg}
\eeq
and writing $\mu(x_2)=t$, (\ref{Int1}) becomes
\beq
v^{(k_1k_2)k_1+k_2}(x_1,z_1,z_2) = h_0 {(cf,df;q)_\infty \over
(g/a,g/b;q)_\infty} \int_{-1}^1 {h(t;g)\over h(t;f)} w(t;a,b,c,d) dt.
\label{Int2}
\eeq
The last integral is a known one; in Ref.~\cite{GR} it is denoted by
$J(a,b,c,d,f,g)$, and shown to be equal to a $_8W_7$ series, i.e.\ a
very-well-poised $_8\vp_7$ series. Using eq.~(6.3.8) of~\cite{GR}, we
obtain 
\bea
&&v^{(k_1k_2)k_1+k_2}(x_1,z_1,z_2)={(ag,bg,cg,df,abcf;q)_\infty \over
(af,bf,g/a,g/b,abcg;q)_\infty} \nn\\
&& \times {}_8W_7(abcgq^{-1};ab,ac,bc,g/d,g/f;q,df).
\label{8W7}
\eea

Let us now consider the other form of
$v^{(k_1k_2)k_1+k_2}(x_1,z_1,z_2)$, given by (\ref{q-v2}). Using
(\ref{q-coup}) and definition (\ref{a-c}), one finds~:
\bea
&&v^{(k_1k_2)k_1+k_2}(x_1,z_1,z_2)= \sum_{n=0}^\infty
{s_n(\mu(x_1);q^{k_1+k_2}s,q^{k_1+k_2}/s |q) \over
\sqrt{(q,q^{2k_1+2k_2};q)_n} } e^{(k_1k_2)k_1+k_2}_n(z_1,z_2) \nn\\
&&=\sum_{n=0}^\infty {(q^{2k_1+2k_2};q)_n\over(q;q)_n}
\left(q^{-2k_1-k_2}z_2/s\right)^n {\;}_3\vp_2\left[
{q^{-n},q^{k_1+k_2}sx_1,q^{k_1+k_2}s/x_1 \atop
q^{2k_1+2k_2},0};q,q\right] \nn\\
&&\qquad\times {\;}_3\vp_2\left[
{q^{-n},q^{2k_1},q^{k_1}z_1/z_2 \atop
q^{2k_1+2k_2},0};q,q\right], 
\label{qv2-real}
\eea
or, putting all this in the notation of (\ref{abcd}) and (\ref{fg}),
\bea
&&v^{(k_1k_2)k_1+k_2}(x_1,z_1,z_2)= \nn\\
&&\quad \sum_{n=0}^\infty
{(abcd;q)_n\over(q;q)_n} \left(f\over abc\right)^n {\;}_3\vp_2\left[
{q^{-n},ac,ad \atop abcd,0};q,q\right] {\;}_3\vp_2\left[
{q^{-n},ab,g/f \atop abcd,0};q,q\right] .
\label{Ser1}
\eea
Equating (\ref{Ser1}) and (\ref{8W7}) gives the $q$-analog of (\ref{ser2}).
Relabelling all the variables, it can be written in the following
form~: 
\begin{prop}
\bea
&&\sum_{n=0}^\infty {\;}_3\vp_2\left[{q^{-n},a,b \atop f,0};q,q\right] 
{\;}_3\vp_2\left[{q^{-n},c,d \atop f,0};q,q\right]
{(f;q)_n\over(q;q)_n} z^n = \nn\\
&&{(abcz,abdz,acdz,bcdz,fz;q)_\infty \over
(acz,bcz,adz,bdz,abcdz;q)_\infty}
{\;}_8W_7(abcdzq^{-1};a,b,c,d,abcdz/f;q,fz) .
\label{Ser2}
\eea
\end{prop}

This formula is the (symmetric) Poisson kernel for Al-Salam--Chihara
polynomials, and was derived earlier by classical methods
in~\cite[(14.8)]{ARS} and~\cite{IS}. As far as we know, this is the
first group theoretical derivation of it.

\setcounter{equation}{0}
\section{Other $su(1,1)$ realizations and their consequences}

In this section two different realizations of $su(1,1)$ and its
positive discrete series representations will be considered. For the
realization considered so far, the typical feature is that the basis
functions $e^{(k)}_n$ are simply monomials $z^n$. Here, the basis
functions will have a more complicated form, and the $q$-analog will
not be treated.

Before turning to the realization, it is useful to observe some
realization-independent facts. In section~II, the operator $X_c$
($0<c<1$) was introduced, and from Table~1 one can see that its
spectrum is identical to that of $J_0$ apart from the factor $(c-1/c)$.
So one can expect that $X_c$ and $J_0$ are related through a unitary
transformation, and this is indeed the case. For $0<c<1$, let $\al>0$ be
defined through
\beq
e^\al={1+c\over 1-c},
\label{alinc}
\eeq
or  inversely,
\beq
c={e^\al-1 \over e^\al+1}.
\label{cinal}
\eeq
Defining, as usual,
\beq
J_1={1\over 2}(J_++J_-),\qquad J_2={1\over 2i}(J_+-J_-)
\eeq
in terms of the $su(1,1)$ basis~(\ref{defsu11}),
it is well known that the following identity
holds~\cite[eq.~(97)]{Adams}~\footnote{This equation can also be
obtained by applying the ``unitary trick'' $J_x \rightarrow iJ_1$, 
$J_y \rightarrow iJ_2$, $J_z \rightarrow J_0$, $\theta \rightarrow
i\alpha$ to the $su(2)$ relation $\exp(-i\th J_y) J_z \exp(i\th J_y) =
(\sin\th) J_x + (\cos\th) J_z$, see~\cite[eq.~(D.10)]{Fano}.}~:
\beq
\exp(i\al J_2) J_0 \exp(-i\al J_2) = (\cosh\al) J_0 - (\sinh\al) J_1.
\label{exp-id}
\eeq
Then it follows from (\ref{Xc}) and the above relation between $c$ and
$\al$ that
\beq
X_c = (c-1/c)\, \exp(i\al J_2) J_0 \exp(-i\al J_2) .
\eeq
Acting with this equation on the formal eigenvectors $v^{(k)}(m)$ of
$X_c$ yields
\beq
J_0 \exp(-i\al J_2) v^{(k)}(m) = (k+m) \exp(-i\al J_2) v^{(k)}(m),
\eeq
thus up to a normalization factor $N_m$ these elements must coincide
with the eigenvectors $e^{(k)}_m$ of $J_0$, leading to
\beq
\exp(i\al J_2) e^{(k)}_m = N_m  v^{(k)}(m).
\label{jg4}
\eeq
In other words, using (\ref{formalv}), Table~1, and the orthogonality
for Meixner polynomials, this yields
\beq
\exp(i\al J_2) e^{(k)}_m = \phi_m \sum_{n=0}^\infty (1-c^2)^k c^{m+n}
\sqrt{(2k)_m (2k)_n\over m!n!} M_n(m;2k;c^2)\, e^{(k)}_n,
\label{jg5}
\eeq
where $\phi_m$ is a phase factor ($|\phi_m|=1$).
This relation is realization-independent, and for particular
realizations it leads to interesting identities. First it should be
noted that for the realization of section~III, the explicit form of
$\exp(i\al J_2) e^{(k)}_m$ becomes rather complicated and (\ref{jg5})
does not reduce to a simple relation. Here, we shall give two different
realizations for which (\ref{jg5}) implies an interesting relation.

First, let $w\geq 1$, $r\in(0,\infty)$ and $p_r=-i\left({d\over dr}
+ {1\over r}\right)$. In~\cite[eq.~(75)]{Adams}, the following
realization of $su(1,1)$ is considered~:
\bea
J_1 &=& {1\over 2} (w^{-2} r^{2-w} p_r^2 +\xi r^{-w}-r^w), \nn\\
J_2 &=& w^{-1} (rp_r-i(w-1)/2), \label{jr6}\\
J_0 &=& {1\over 2} (w^{-2} r^{2-w} p_r^2 +\xi r^{-w}+r^w). \nn
\eea
These provide a realization of the $su(1,1)$ basis which are Hermitian
under the scalar product
\beq
\langle f, g \rangle = \int_0^\infty f^*(r) g(r) r^w dr.
\eeq
More precisely, ${\cal D}^+(k)$ consist of functions of $r$ that are
integrable with respect to the above measure, and (\ref{jr6}) forms a
realization of the $su(1,1)$ basis in this representation ${\cal
D}^+(k)$ provided
\beq
\xi = k(k-1)-W(W-1),\qquad\hbox{where}\qquad W={w+1\over 2w}.
\eeq
The normalized eigenfunctions $e^{(k)}_m$ of $J_0$ are given by 
\beq
e^{(k)}_m = 2^W \sqrt{w\,m! \over\Gamma(2k+m)} \exp(-r^w) \left( 2r^w
\right)^{k-W} L_m^{(2k-1)}(2r^w),
\label{jg9}
\eeq
in terms of the Laguerre polynomials~(\ref{Laguerre}),
see~\cite[eq.~(89)]{Adams}. 
An advantage of this realization is that $J_2$ is a linear combination
of $r{d\over dr}$, and hence $\exp(i\al\,J_2)$ acting on a function of
$r$ takes a simple form, i.e.
\beq
\exp(i\al\,J_2) f(r) = e^{\al W} f(e^{\al/w}r).
\label{jg8}
\eeq
Using all this information, (\ref{jg5}) becomes
\bea
&&{m!\over (2k)_m} \exp(-e^\al r^w)\, e^{\al k}
L_m^{(2k-1)}(2 e^\al r^w) = \\
&&\phi_m \sum_{n=0}^\infty (1-c^2)^k\, c^{n+m} M_n(m;2k;c^2) \exp(-r^w)
L_n^{(2k-1)}(2 r^w).
\label{jg5b}
\eea
Making the replacements
\beq
x=2r^w,\qquad \rho=e^\al >1, \qquad a=2k-1,
\eeq
and keeping in mind the relation~(\ref{cinal}), (\ref{jg5b}) takes a
simpler form. In fact, rather than leaving this as an infinite series
expression in terms of Laguerre polynomials, it is more convenient to
integrate with respect to the orthogonality measure of Laguerre
polynomials and then obtain the equivalent expression~:
\begin{coro}
For $\rho>1$ and $a>-1$,
\bea
&&\int_0^\infty L_m^{(a)}(\rho x) L_n^{(a)}(x) \exp(-(\rho+1)x/2) x^a dx =
\label{jg5c} \\
&& (-1)^m {\Gamma(a+n+1)\over n!} {(a+1)_m\over m!} \left(2\over
\rho+1\right)^{a+1} \left(\rho-1\over\rho+1\right)^{n+m} M_n(m;a+1;
\left(\rho-1\over\rho+1\right)^2). \nn
\eea
\end{coro}
The fact that the phase factor $\phi_m$ is equal to $(-1)^m$ can easily
be derived from the case $n=0$. 
Although we have not found~(\ref{jg5c}) in the literature, it is
perhaps not new, and it can probably be derived by classical methods as
well. Nevertheless, in our treatment it does not require any extra work
to derive it, and the appearance of a Meixner polynomial in the rhs
of~(\ref{jg5c}) has a natural explanation.

As a second example of a different realization, we shall consider the
well known boson realization of $su(1,1)$, see also~\cite{Adams}. In
terms of the annihilation and creation operators
\beq
a ={1\over \sqrt{2}}\left(x+{d\over dx}\right),\qquad
a^\dagger ={1\over \sqrt{2}}\left(x-{d\over dx}\right),
\eeq
we have the realization
\bea
J_1 &=& {1\over 4} \Bigl( (a^\dagger)^2+ a^2 \Bigr), \nn\\
J_2 &=& {1\over 4i} \Bigl( (a^\dagger)^2- a^2 \Bigr),\\
J_0 &=& {1\over 4} \Bigl( a a^\dagger+ a^\dagger a \Bigr). \nn
\eea
The representation space is $L_2(\Real)$, with inner product $\langle
f,g \rangle = \int f^*(x) g(x) dx$. This corresponds to the positive
discrete series representation ${\cal D}^+(1/4)$, so $k=1/4$. The
normalized eigenvectors $e^{(1/4)}_n$ are given by
\beq
e^{(1/4)}_n = {1\over \pi^{1/4} 2^n \sqrt{(2n)!}} \exp(-x^2/2)
H_{2n}(x), 
\eeq
with $H_n(x)$ the usual notation for Hermite polynomials~\cite[\S
22]{Abra}. 
Since also for this realization $J_2$ is a linear combination of
$x{d\over dx}$, the simple relation
\beq
\exp(i\al J_2) f(x) = e^{-\al/4} f(e^{-\al/2} x)
\eeq
holds.
Plugging all this information in (\ref{jg5}) gives an expression
similar to (\ref{jg5b}), but with Hermite polynomials instead of
Laguerre polynomials. Replacing herein $e^{-\al/2}$ by $\la$,
and going to the equivalent form in terms of an integral, this result
can be expressed as follows~:
\begin{coro}
\bea
&&\int_{-\infty}^\infty H_{2m}(\la x) H_{2n}(x) \exp(-(\la^2+1)x^2/2) dx =
\label{jg5d} \\
&& (-1)^m \sqrt{2\pi\over 1+\la^2}
\left(1-\la^2\over1+\la^2\right)^{n+m} {(2m)!(2n)!\over m!n!} M_n(m;1/2;
\left(1-\la^2\over1+\la^2\right)^2).
\nn
\eea
\end{coro}
The phase factor $\phi_m$ is again easily derived.
Again we have not found~(\ref{jg5d}) in the literature. Although it can
be derived by classical methods, here it has required no extra work.

With the two different realizations considered in this section, the
analysis can be continued. For example, another interesting equation is
the realization of~(\ref{vv}) in terms of the above Laguerre
polynomials. 

In the case of the quantum algebra $U_q(su(1,1))$, similar developments
can be made provided one can find sufficiently simple realizations in
terms of $q$-difference operators and $q$-special functions.

To conclude, using the explicit knowledge of the expansion coefficients
of the eigenstates of a general Hamiltonian in terms of the $J_0$
eigenstates, we have deduced a number of identities for orthogonal
polynomials. By choosing a realization for which the $J_0$ eigenstates
coincide with monomials $z^n$, generating functions or Poisson kernels
were derived. Considering the $q$-analog of this realization, the Poisson
kernel for Al-Salam--Chihara polynomials is found. Finally, we have
shown that different realizations for the $su(1,1)$ case give rise
to explicit formulas for integrals over orthogonal polynomials
expressed as a Meixner polynomial. 

\section*{Acknowledgements}

It is a pleasure to thank Prof.\ K.\ Srinivasa Rao for stimulating
discussions. 
This research was partly supported by the E.C. (contract No.
CI1*-CT92-0101).

\end{document}